\documentclass{article}
\usepackage{icrctc07}

\title{AMIGA, Auger Muons and Infill for the Ground Array}
\shorttitle{AMIGA}
\authors{A. Etchegoyen$^{1}$, for the Pierre Auger Collaboration$^{2}$.}
\shortauthors{A. Etchegoyen for the Pierre Auger Collaboration}
\afiliations{$^{1}$ Departamento de F\'isica (Tandar), Centro
At\'omico Constituyentes, Comisi\'on Nacional de Energ\'ia
At\'omica and UTN-FRBA, \\ $^{2}$ Observatorio Pierre Auger, Av.
San Mart\'in Norte 304 (5613) Malarg\"{u}e, Prov. Mendoza,
Argentina} \email{etchegoy@tandar.cnea.gov.ar}

\abstract{The Pierre Auger Observatory is planned to be upgraded so
that the energy spectrum of cosmic rays can be studied down to 0.1
EeV and the muon component of showers can be determined.  The former
will lead to a spectrum measured by one technique from 0.1 EeV to
beyond 100 EeV while the latter will aid identification of the
primary particles. These enhancements consist of three high
elevation telescopes (HEAT) and an infilled area having both surface
detectors and underground muon counters (AMIGA). The surface array
of the Auger Observatory will be enhanced over a 23.5 km$^{2}$ area
by 85 detector pairs laid out as a graded array of water-Cherenkov
detectors and 30 m$^{2}$ buried muon scintillator counters.  The
spacings in the array will be 433 and 750 m. The muon detectors will
comprise highly segmented scintillators with optical fibres ending
on multi-anode phototubes. The AMIGA complex will be centred 6.0 km
away from the fluorescence detector installation at Coihueco and
will be overlooked by the HEAT telescopes. We describe the design
features of the AMIGA enhancement.}

\email{etchegoy@tandar.cnea.gov.ar}

\begin{document}
\maketitle

The cosmic ray spectrum shows three features at higher energies: the
second knee, the ankle, and the GZK-cut off, and in order to
seamless study this region \cite{GMT07} Auger will be upgraded with
HEAT (High Elevation Auger Telescopes, \cite{HEAT07}) and AMIGA
(Auger Muons and Infill for the Ground Array). These two
enhancements will encompass the second knee - ankle region where the
transition from galactic to extra galactic cosmic rays is assumed to
occur. The two main experimental requirements are good energy
resolution in order to obtain the spectrum and primary type
identification since the galactic (heavy primaries) to extra
galactic (light primaries) source transition is directly linked to
primary composition.

\begin{figure}
\begin{center}
\includegraphics [width=0.48\textwidth]{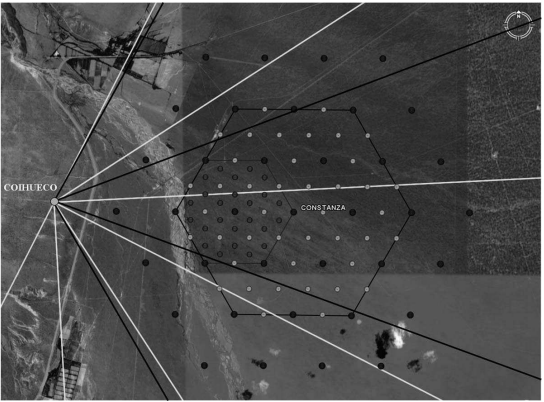}
\end{center}
\caption{Layout of Auger enhancements. White and black lines show
the six original and three enhanced telescopes FOVs, respectively.
Grey, white and black dots indicate SDs plus buried muon counters
placed 433, 750, and 1500 m apart, respectively. In this area a
further enhancement of radio detection of extensive air showers
will start its R\&D phase \cite{ADB07}.}\label{fig:layout}
\end{figure}

In this note we concentrate on AMIGA. It will consist of 85 pair
of water Cherenkov surface detectors (SD) and 30 m$^{2}$ plastic
scintillators buried $\sim$ 3.0 m underground, placed in a graded
infill of 433 and 750 m triangular grids (see Fig.
\ref{fig:layout}) overlooked by the fluorescence detectors (FD)
(including HEAT) placed at the Coihueco hill top. AMIGA graded
infilled areas are bound by the two hexagons shown in the figure
covering areas of 5.9 and 23.5 km$^{2}$.

In regards to the mentioned spectrum measurement with good energy
resolution, Auger hybrid detecting system was conceived in order
to perform careful systematic uncertainty crosschecks which are
currently under way. They will eventually permit to consolidate an
energy spectrum of unprecedented precision.  Main uncertainties in
the energy calibration are the absolute calibration, the
atmospheric light attenuation, and the fluorescence yield for the
FD system, and the simulated airshower muon component for the SD
system. There are clear indications that simulations under predict
the shower muon contents \cite{ENG07} and as such the Auger SD
energy estimator will be biased and this bias would increase with
zenith angle \cite{BIL07}. Large muon counters will aid towards
solving this problem by directly measuring the number of muons
with reduced poisson fluctuations.

In regards to composition analyses, the two relevant shower
parameters are the atmospheric depth at shower maximum, $X_{max}$,
and the shower muon contents. Other composition sensitive
parameters dependent on them. Gamma-hadron discrimination is
easier to perform than hadron-hadron discrimination since at E
$\geq$ 0.1 EeV, $X_{max}$ values for gamma induced showers are
already well above those from hadron primary showers \cite{RIS07}.
Also gamma showers are essentially electromagnetic with a
vanishing muon component. No photon detection has been reported so
far and a direct detection at E $\geq$ 1.0 EeV will encourage
theoretical models such as correlations with nuclear primaries
\cite{KUS06} and with super heavy dark matter decays (\cite{RIS07}
and references within). In this latter supposition, an E $\geq$
0.1 EeV gamma flux enhanced from the galactic center might be
detected without a higher-energy counterpart. A further advantage
of photon detection is that their sources are easier to identify
since photons are not deflected by electromagnetic fields.

Composition is very poorly understood in this energy range where a
wide variety of mixed compositions are reported ranging from
proton to iron dominated primaries (\cite{ANC06} and references
within). Still, hadron composition can only be assessed within a
given hadronic interaction model and much more robust results are
attained from the variation rate of either $X_{max}$ (the
elongation rate) or muon contents as a function of energy
\cite{ABU00}. A simultaneous change detected by both FD and muon
counters will be the most compelling evidence of a composition
change casting light on the transition of cosmic ray sources from
galactic to extra galactic origins \cite{ETC07}.

AMIGA reconstruction performances are quite encouraging, they have
been outlined in \cite{MED06, ETC07} for tank infilled areas and
muon counters, respectively. Suffice to say that the surface
detector reconstruction is currently well understood by the Auger
collaboration and that we have developed \cite{SUP07} a detailed
muon reconstruction system which is based on the parameterized
muon lateral distribution function \cite{BUR05} currently used by
KASCADE-Grande. The scintillator modules are simulated and the
reconstruction procedure includes saturated (more than 90 muons)
and silent (0,1, or 2 muons) counters. The shower reconstructed
parameter is $N_{\mu}(600)$, the estimated number of muons 600 m
away from the shower axis, an excellent primary type indicator.

\begin{figure}
\begin{center}
\includegraphics [width=0.48\textwidth]{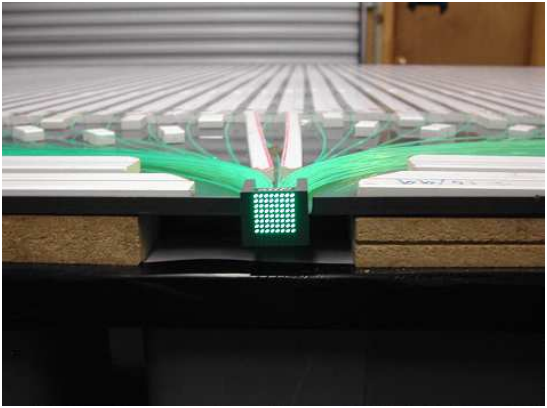}
\end{center}
\caption{Muon counter assembly at Argonne National Laboratory of a
64 200 cm long prototype displaying the 64 pixel optical
connector. The 4.1 cm wide strips and the green fibers are also
shown.}\label{fig:counter}
\end{figure}

In this note we are concentrating in the muon detector hardware.
These counters will comprise highly segmented scintillators (to
avoid under counting) with optical fibres ending on 64-pixel
multi-anode photo multiplier tubes (PMT). The design adopts
similar scintillator strips as for the MINOS experiment
\cite{MIN98}. The current baseline design calls for 400 cm long
$\times$ 4.1 cm wide $\times$ 1.0 cm high strips of extruded
polystyrene doped with fluors and co-extruded with TiO$_{2}$
reflecting coating with a groove in where a wavelength shifter
fibre is glued (see Fig. \ref{fig:counter}) and covered with
reflective foil. Each module will consist of 64 strips with the
fibres ending on an optical connector matched to a 64 multianode
Hamamatsu H7546B PMT of 2mm $\times$ 2mm pixel size lodged in a
PVC casing. Each muon counter will be composed of three of these
modules buried alongside a water Cherenkov tank, i.e. 192
independent channels.

The response of each scintillator strip will be characterized
using a 5 mCu $^{137}$Cs radioactive source mounted on a scanner
designed for this purpose. The scanner is an X-Y positioning
system with four tooth belt activated linear guides moved by two
step-by-step motors of 8.7 Nm torque. The whole positioning system
has up to 1 mm precision in any of the two axis and an effective
total displacement of 5 m $\times$ 3.75 m. It will be mounted on a
structure built of Rexroth (Bosch group) aluminum profiles which
supports both the modules and the X-Y positioning system. This
method has been validated by measuring ratios of collected charges
and ratios of currents from scintillators strips with different
optical fibers. The experimental layout consisted of three
scintillator strips, each one with different wave length shifter
optical fibers (Kuraray 1.0 mm, Kuraray 1.2 mm, and Bycron 1.2 mm)
channelled to different pixels of a multianode PMT. Charges were
collected from background muons impinging on the scintillators
while currents were produced via a 20 $\mu$Cu $^{137}$Cs source
mounted on top of the scintillators.

AMIGA electronics will have both an underground and a surface
section powered by solar panels. Each of the three underground
modules per counter will have attached a PCB with a data handling
FPGA and a communication and monitoring system with a
microcontroller. Each electronic channel will have an amplifier
and a discriminator, set at $\sim$ 30\% of the pixel mean single
photo electron amplitude. The pulses from all strips are
synchronized to the 40 MHz water tank clock and after
discrimination each strip output will either be ´0´ or ´1´. These
numbers are stored in a circular buffer and upon reception of a
tank trigger signal, they are adequately channelled by the
microcontrollers to the surface electronics. On the surface, a
second data handling FPGA receives the data and sends it to a
microcomputer, which upon a request from the central data
acquisition system at the Auger Campus, transmits both tank and
muon counter data by radio link. Also, and in order to test each
strip, a monitoring mode is envisaged by detecting atmospheric
background muons.

\begin{figure}
\begin{center}
\includegraphics [width=0.48\textwidth]{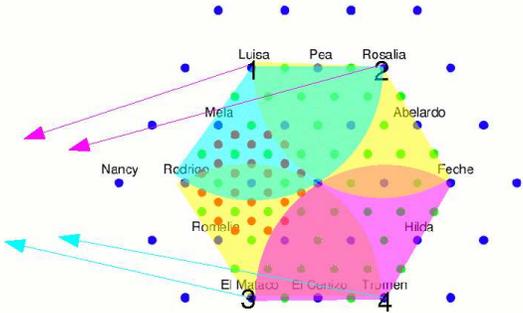}
\end{center}
\caption{AMIGA two level telecommunication star topology.
Concentrators are labelled from 1 to 4. In this layout
concentrator-subscriber systems 2 and 3 use the same 802.11
independent channel. Bottom and upper arrows show channelling of
data to the two access points.}\label{fig:comms}
\end{figure}

AMIGA will communicate using an 802.11 standard wireless network,
widely known as WiFi, taking advantage of the low cost and wide
availability of this technology. As a plus, the high communication
bandwidth will allow for the recollection of high amount of data
at the early stages of the experiment, in order to study better
the characteristics of muon counters and showers. This system must
also be capable of carrying the surface detectors data and so the
wireless local area network must integrate with the existing
communications. The antennas and physical network topology are
carefully chosen to cover long distances and avoid interferences.
A two level star topology was chosen by setting concentrators
relatively close to the subscriber stations. These concentrators
are the center of four lower level logical stars (see Fig.
\ref{fig:comms}) and they have a directional 120$^{\circ}$ sector
antenna which collects the signal from the stations directional
antennas. These stations use three 802.11 independent channels but
since there are four concentrators, one of these channels is used
twice.

The higher level star is formed by two access points (AP) located
at the Coihueco fluorescence observatory tower which collect the
signal from the concentrators. Each AP will use an 802.11
independent channel, with two concentrators communicating to each
channel. To minimize interference, the higher and lower levels
stars will use different polarizations. Careful selection of the
cell distribution and long range link paths must be made to
further reduce interference.

An estimation of the number of events per year with energy larger
than $E_{0}$ and zenith angle below $\theta_{max} = 60^{\circ}$
was obtained by assuming a cosmic ray flux following a power law
with spectral index -2.84 as quoted by Auger (\cite{MED06} and
references within), is displayed in Table \ref{tab:flux}.

\begin {table}
\begin{tabular}{|c|c|c|}
  \hline
  $E_{o}$ [EeV] & Area [km$^{2}$] & No. events year$^{-1}$ \\
  \hline
  0.1 & 5.9 & 16000 \\
  0.3 & 23.5 & 8500 \\
  \hline
\end{tabular}
\caption {Expected number of events per year with the AMIGA
infilled areas.}\label{tab:flux}
\end {table}

AMIGA will start by deploying a prototype, after full
commissioning of the 1500 m grid array which is planned to occur
early 2008. This prototype will permit to gain  experience on muon
counters and experimentally estimate possible punch-throughs.  It
is designed to have three 4 m$^{2}$ X-Y parallel plates buried at
three different depth, near the surface, at $\sim$ 3.0 m deep, and
in between (this configuration might also shed some light on the
muon momentum spectrum \cite{BIL07}).  A unitary cell of seven
detector pairs will be also deployed.  This unitary cell will
consist in detector pairs deployed $\sim$ 3.0 m underground at
each vertex of a regular 750 m hexagon. These two prototype
systems will be operated for one year prior to full deployment of
the 750 m infill muon counters planned to start by mid 2009.  The
750 infilled tank array will begin deployment simultaneously with
the muon prototypes since this technology has been fully tested in
Auger. The 433 m infill will start after completion of the 750 m
infill.

\end{document}